\documentclass{pspum-l}

\providecommand{\ch}{\textnormal{ch}}
\providecommand{\Tr}{\textnormal{Tr}}
\providecommand{\ev}{\textnormal{ev}}
\providecommand{\pfaff}{\textnormal{pfaff}}
\providecommand{\Hom}{\textnormal{Hom}}
\providecommand{\Hol}{\textnormal{Hol}}

\theoremstyle{definition}

\theoremstyle{remark}

\numberwithin{equation}{section}

\begin{document}

\title{Freed-Witten anomaly and D-brane gauge theories}

\author{Fabio Ferrari Ruffino}
\address{ICMC - Universidade de S\~ao Paulo, Avenida Trabalhador s\~ao-carlense 400, 13566-590 - S\~ao Carlos - SP, Brasil}
\email{ferrariruffino@gmail.com}
\thanks{The author was supported by FAPESP (Funda\c{c}\~ao de Amparo \`a Pesquisa do Estado de S\~ao Paulo).}

\date{October 25, 2011.}

\keywords{Algebraic and differential topology, D-brane gauge theory}

\begin{abstract}
We discuss the different nature of the gauge theories on a D-brane or a stack of D-branes, in type II superstring theory, as follows from the Freed-Witten anomaly. Usually on a D-brane world-volume there is a standard gauge theory, described by the A-field thought of as a connection on a complex vector bundle. Actually, this is a particular case, even if it is the most common one. In order to get a complete picture, within the framework provided by the geometry of gerbes with connection, it is necessary to give a joint geometrical description of the A-field and the B-field, via the language of $\rm\check{C}$ech hypercohomology. The Freed-Witten anomaly, which is a global world-sheet anomaly, imposes some constraints on such fields: we will show for each case what is the nature of the corresponding gauge theory on the D-brane or stack of D-branes.
\end{abstract}

\maketitle

\section{Introduction}

In the papers \cite{BFS, FR3} we have shown how to classify the admissible configurations of the $A$-field and the $B$-field, in the presence of D-branes that are free from the Freed-Witten anomaly. We worked in the framework of type II superstring theory, and the mathematical language we used is the geometry of gerbes with connection, described via the $\rm\check{C}$ech hypercohomology of sheaves. In this paper we summarize how this classification determines the nature of the gauge theory on the world-volume, starting from the definition of ``non-integral vector bundles'', on which the gauge theory is defined when the B-field is flat. The paper is organized as follows. In section \ref{NonIntegralVB} we review the definition of non-integral vector bundles, starting from the case of line bundles. In section \ref{GaugeDBrane} we show what are the possible geometrical natures of the gauge theory on a world-volume, within this framework. In section \ref{Residual} we describe a residual gauge freedom depending on the topology of the space-time, and we show that it has a link with the twisted K-theory classification of D-brane charges. In section \ref{Conclusions} we draw our conclusions.

\section{Non-integral vector bundles}\label{NonIntegralVB}

\subsection{Line bundles}

Let us consider a smooth manifold $X$ with a good cover $\mathfrak{U} = \{U_{i}\}_{i \in I}$. The group of isomorphism classes of complex line bundles (with hermitian metric) is isomorphic to the $\rm\check{C}$ech cohomology group $\check{H}^{1}(X, \underline{U}(1))$, for $\underline{U}(1)$ the sheaf of $U(1)$-valued smooth functions on $X$. Via the exact sequence of sheaves:
\begin{equation}\label{ExSeqZRU1}
	0 \longrightarrow \mathbb{Z} \longrightarrow \underline{\mathbb{R}} \overset{e^{2\pi i \cdot}}\longrightarrow \underline{U}(1) \longrightarrow 0
\end{equation}
(where $\underline{\mathbb{R}}$ is the sheaf of real smooth functions on $X$) we get an isomorphism $\check{H}^{1}(X, \underline{U}(1)) \simeq H^{2}(X, \mathbb{Z})$. Therefore, a line bundle is completely determined up to isomorphism by its corresponding class in $H^{2}(X, \mathbb{Z})$, called first Chern class. If we consider the group of line bundles with connection, it is isomorphic to the $\rm\check{C}$ech hypercohomology group \cite{Brylinski}:
\begin{equation}\label{H1Hyper}
	\check{H}^{1}(X, \underline{U}(1) \overset{\tilde{d}}\longrightarrow \Omega^{1}_{\mathbb{R}})
\end{equation}
for $\Omega^{1}_{\mathbb{R}}$ the sheaf of real-valued differential forms of degree 1 on $X$, and $\tilde{d} := \frac{1}{2\pi i} d \circ \log$. Taking the exterior differential of the connection 1-forms we can define the curvature, which is a global integral (up to a multiplicative constant) 2-form $F$. If $(L, \nabla)$ is a line bundle with connection, $c_{1}(L)$ the first Chern class and $F$ the curvature, then $[\frac{i}{2\pi}F]_{dR} \simeq c_{1}(L) \otimes_{\mathbb{Z}} \mathbb{R}$, where $[F]_{dR}$ is the de-Rham cohomology class of $F$.

We can consider more generally a ``non-integral line bundle'', which is represented by transition functions $g_{\alpha\beta}: U_{\alpha\beta} \rightarrow U(1)$ such that $g_{\alpha\beta}g_{\beta\gamma}g_{\gamma\alpha}$, instead of being $1$, is only required to be locally constant. Two set of transition functions are still considered equivalent when they differ by a coboundary of $\underline{U}(1)$. Such transition functions form a cocycle in the $\rm\check{C}$ech cochain complex of the quotient sheaf $\underline{U}(1)/U(1)$, where $U(1)$ denotes the \emph{constant} sheaf. Hence, we can generalize the exact sequence \eqref{ExSeqZRU1} to:
\begin{equation}\label{ExSeqZRU1Const}
	0 \longrightarrow \mathbb{R} \longrightarrow \underline{\mathbb{R}} \overset{e^{2\pi i \cdot}}\longrightarrow \underline{U}(1)/U(1) \longrightarrow 0.
\end{equation}
Since a non-integral bundle projects to a class in $\check{H}^{1}(X, \underline{U}(1)/U(1))$, via the Bockstein map of \eqref{ExSeqZRU1Const} we can associate to it a first Chern class in $H^{2}(X, \mathbb{R})$, which is therefore not necessarily integral. We can endow a non-integral line bundle with a connection, getting a class in the hypercohomology group:
\begin{equation}\label{H1HyperConst}
	\check{H}^{1}(X, \underline{U}(1)/U(1) \overset{\tilde{d}}\longrightarrow \Omega^{1}_{\mathbb{R}}).
\end{equation}
We define the curvature in the same way, and we get a global 2-form $F$ which is not necessarily integral any more, since its de-Rham cohomology class still corresponds to the (real) first Chern class of the non-integral line bundle.

\subsection{Vector bundles} We can now define ``non-integral bundles'' of any rank. A rank-$n$ vector bundle (with hermitian metric) is represented by transition functions $g_{\alpha\beta}: U_{\alpha\beta} \rightarrow U(n)$, satisfying the cocycle condition. If we endow it with a connection, it is represented by the transition functions and the connection 1-forms $A_{\alpha}: TU_{\alpha} \rightarrow i\mathfrak{u}(n)$, for $\mathfrak{u}(n)$ the Lie algebra of $U(n)$, such that $A_{\beta} - g_{\alpha\beta}^{-1}A_{\alpha}g_{\alpha\beta} = \frac{1}{2\pi i} g_{\alpha\beta}^{-1}dg_{\alpha\beta}$. The curvature is a local form $F_{\alpha} = dA_{\alpha} + A_{\alpha} \wedge A_{\alpha}$, whose gauge-transformations are the action of the transition functions by coniugation. Hence we can define the Chern classes via the symmetric polynomials $P_{i}$, which are invariant by coniugation, as:
\begin{equation}\label{ChernClasses}
	c_{i}[\{g_{\alpha\beta}, A_{\alpha}\}] = [P_{i}(\textstyle \frac{i}{2\pi} \displaystyle F)].
\end{equation}
As above, we can define a non-integral rank-$n$ vector bundle as a class of transition functions $g_{\alpha\beta}: U_{\alpha\beta} \rightarrow U(n)$ such that $g_{\alpha\beta}g_{\beta\gamma}g_{\gamma\alpha}$, instead of being 1, is locally constant, and its value on a triple intersection belongs to the center of $U(n)$, i.e.\ $U(1) \cdot I_{n}$ for $I_{n}$ the identity matrix. This is a particular case of twisted vector bundle \cite{Karoubi}, since the latter is represented by transition functions such that $g_{\alpha\beta}g_{\beta\gamma}g_{\gamma\alpha}$ takes values in the center of $U(n)$, without being locally constant in general. We can endow a non-integral vector bundle with a connection, which is represented by 1-forms $A_{\alpha}: TU_{\alpha} \rightarrow i\mathfrak{u}(n)$ satisfying, as in the ordinary case, $A_{\beta} - g_{\alpha\beta}^{-1}A_{\alpha}g_{\alpha\beta} = \frac{1}{2\pi i} g_{\alpha\beta}^{-1}dg_{\alpha\beta}$. It is easy to show that the gauge transformations of the curvature are still the action by coniugation of the transition functions, contrary to the general case of twisted bundles, so that equations \eqref{ChernClasses} still gives well-defined cohomology classes in $H^{\ev}(X, \mathbb{R})$, which depends only on the bundle (not on the connection), but they are not necessarily integral. Thus, even the Chern character, defined as usual as:
	\[\ch[\{g_{\alpha\beta}, A_{\alpha}\}] = [\Tr\,\exp(\textstyle \frac{i}{2\pi} \displaystyle F)],
\]
in the case of non-integral bundles takes values in a lattice which is different from the one of ordinary vector bundles.

\section{Gauge theories on a D-brane}\label{GaugeDBrane}

We now show how non-integral bundles are related to the gauge theory on a D-brane or stack of D-branes world-volume, in the framework of type II superstring theory with non-trivial $B$-field in general.

\subsection{Single D-brane}

We recall that, for a single D-brane, the world-sheet path-integral measure contains the terms \cite{FW}:
\begin{equation}\label{PathIntegralDBrane}
	e^{iS} = \cdots \pfaff \, D_{\phi} \cdot \exp\biggl(\,2\pi i \cdot \int_{\Sigma} \phi^{*}B\,\biggr) \cdot \exp\biggl(\, 2\pi i \cdot \int_{\partial \Sigma} \phi^{*}A \,\biggr)
\end{equation}
where $\Sigma$ is the world-sheet, $\phi$ is the trajectory of the world-sheet in the space-time, $\pfaff \, D_{\phi}$ is the pfaffian of the Dirac operator on the world-sheet, coupled to the space-time via $\phi$, and $A$ and $B$ are the $A$-field and the $B$-field. The term involving the $B$-field can be interpreted as the holonomy over $\Sigma$ of a gerbe with connection \cite{Brylinski}. We recall that the latter is an element of the hypercohomology group: \begin{equation}\label{H2Hyper}
	\check{H}^{2}(X, \underline{U}(1) \overset{\tilde{d}}\longrightarrow \Omega^{1}_{\mathbb{R}} \overset{d}\longrightarrow \Omega^{2}_{\mathbb{R}}).
\end{equation}
In this case, $X$ is the space-time, the connection 2-forms are the local representatives of the $B$-field, while the curvature, which is a global 3-form, is the $H$-flux. Actually the $B$-field should be described in a more refined way \cite{DFM, DFM2}, but this is enough for our purposes. Similarly, the pfaffian $\pfaff \, D_{\phi}$ is a section of a line bundle over the loop space of $Y$, which is determined \cite{Brylinski} by a \emph{flat} gerbe with connection on the world-volume $Y$, called the \emph{spin$^{c}$-gerbe}, whose first Chern class is the third integral Stiefel-Whitney class $W_{3}(Y) \in H^{3}(Y, \mathbb{Z})$, and whose flat holonomy is the second Stiefel-Whitney class $w_{2}(Y) \in H^{2}(Y, \mathbb{Z}_{2})$ \cite{LM}. Since $\Sigma$ is an open surface, such that $\phi(\partial \Sigma) \subset Y$, the product of the two terms:
	\[\pfaff \, D_{\phi} \cdot \exp\biggl(\,2\pi i \int_{\Sigma} \phi^{*}B\,\biggr)
\]
is in general a section of a line bundle over the space of maps from $\Sigma$ to $X$, sending $\partial \Sigma$ to $Y$, and becomes a well-defined number only when the tensor product of the two gerbes on $Y$ is trivial and trivialized. If we call $[H] \in H^{3}(X, \mathbb{Z})$ the integral first Chern class of the $B$-field gerbe, it must therefore happen that:
\begin{equation}\label{FWDBrane}
	W_{3}(Y) + [H]\vert_{Y} = 0
\end{equation}
and the term $W_{3}(Y) + [H]\vert_{Y}$ is called \emph{Freed-Witten anomaly} \cite{FW}. In this case, the $A$-field provides a reparametrization of the tensor product of the two gerbes, which trivializes it. Therefore, if we represent the $B$-field gerbe, restricted to $Y$, as $\{g_{\alpha\beta\gamma}, -\Lambda_{\alpha\beta}, B_{\alpha}\}$ and the spin$^{c}$ gerbe as $\{\eta_{\alpha\beta\gamma}^{-1}, 0, 0\}$, for $\eta_{\alpha\beta\gamma}^{-1}$ representing $w_{2}(Y)$ in the cohomology of the constant sheaf $U(1)$, then it must hold that:
	\[\{g_{\alpha\beta\gamma}\eta_{\alpha\beta\gamma}^{-1}, -\Lambda_{\alpha\beta}, B_{\alpha}\} \cdot \check{\delta}^{1}\{h_{\alpha\beta}, A_{\alpha}\} = \{1, 0, B + F\}
\]
i.e.:
\begin{equation}\label{AFieldDBrane}
	\check{\delta}^{1}\{h_{\alpha\beta}, A_{\alpha}\} = \{g_{\alpha\beta\gamma}^{-1}\eta_{\alpha\beta\gamma}, \Lambda_{\alpha\beta}, B + F\}
\end{equation}
for $F\vert_{U_{\alpha}} = dA_{\alpha}$. Since the choice of the representatives is arbitrary, we see with the language of $\rm\check{C}$ech hypercohomology the well-known fact that, in general, $B$ and $F$ are not gauge-invariant, because, even if they are chosen to be globally defined, they are subject to large gauge transformations $B \rightarrow B + \Phi$ and $F \rightarrow F - \Phi$. The situation is different when the $H$-flux, restricted to the world-volume, is zero, i.e.\ when the $B$-field gerbe is flat on the world-volume. In this case, we can choose a cocycle $\{g_{\alpha\beta\gamma}, 0, 0\}$ for the $B$-field gerbe on the world-volume: this is clear by analogy with flat line bundles, for which, choosing \emph{parallel} sections, the transition functions become locally constant and the connection 1-forms vanish. In this case, \eqref{AFieldDBrane} becomes:
\begin{equation}\label{AFieldDBraneFlat}
	\check{\delta}^{1}\{h_{\alpha\beta}, A_{\alpha}\} = \{g_{\alpha\beta\gamma}^{-1}\eta_{\alpha\beta\gamma}, 0, F\}
\end{equation}
which is the equation of a \emph{non-integral} line bundle. That's why, in general, when the large gauge transformations can be canonically fixed, we do not always get a gauge theory in the usual sense, but a gauge theory on a non-integral line bundle. In particular, we miss the information about the torsion part (e.g.\ the Aharonov-Bohm effect in electromagnetism), and the Wilson loop of the $A$-field is a section of a flat line bundle over the loop space on $Y$, rather than a well-defined complex number of modulus $1$, since it compensates in \eqref{PathIntegralDBrane} the anomaly of the other two terms.

If it turns out that $w_{2}(Y) = 0$ and that the $B$-field gerbe, restricted to $Y$, has trivial holonomy, than we can choose transition functions equal to $1$ for both the gerbes, so that \eqref{AFieldDBraneFlat} becomes:
\begin{equation}\label{AFieldDBraneStandardGT}
	\check{\delta}^{1}\{h_{\alpha\beta}, A_{\alpha}\} = \{1, 0, F\}.
\end{equation}
Only in this case, which is of course the most common one, we get a canonical gauge theory, in the sense that it is defined on a line bundle in the usual sense, so that we recover the information even about the torsion part, and we can define the holonomy of the $A$-field over a loop as a number.

\subsection{Stack of D-branes}

For a stack of D-branes the action \eqref{PathIntegralDBrane} becomes \cite{Kapustin}:
\begin{equation}\label{PathIntegralStack}
	e^{iS} = \cdots \pfaff \, D_{\phi} \cdot \exp\biggl(\,2\pi i \int_{\Sigma} \phi^{*}B\,\biggr) \cdot \Tr \, \mathcal{P} \exp\biggl(\, 2\pi i \int_{\partial \Sigma} \phi^{*}A \,\biggr)
\end{equation}
where $\mathcal{P}$ is the path-ordering operator. The terms involving the $B$-field and the pfaffian of the Dirac operator remains unchanged, therefore we still need to trivialize the product of the spin$^{c}$-gerbe and the $B$-field gerbe restricted to the D-brane. Thus we still need that:
\begin{equation}\label{AFieldStack}
	\check{\delta}^{1}\{h_{\alpha\beta}, A_{\alpha}\} = \{g_{\alpha\beta\gamma}^{-1}\eta_{\alpha\beta\gamma}, \Lambda_{\alpha\beta}, B + \textstyle \frac{1}{n} \displaystyle \Tr F\},
\end{equation}
but this time the coboundary $\check{\delta}^{1}$ has a different meaning, since the transition functions $h_{\alpha\beta}$ take values in $U(n)$, thus they are not necessarily a trivialization of $g_{\alpha\beta\gamma}^{-1}\eta_{\alpha\beta\gamma}$ in the $\underline{U}(1)$-cohomology. They define a twisted vector bundle. Anyhow, if $h_{\alpha\beta}h_{\beta\gamma}h_{\gamma\alpha} = h_{\alpha\beta\gamma} \cdot I_{n}$, then the class $[\{h_{\alpha\beta\gamma}\}] \in H^{2}(Y, \underline{U}(1)) \simeq H^{3}(Y, \mathbb{Z})$ must be a \emph{torsion class}, because, taking the determinants, we get $\det h_{\alpha\beta} \cdot \det h_{\beta\gamma} \cdot \det h_{\gamma\alpha} = (h_{\alpha\beta\gamma})^{n}$, and, since $\det h_{\alpha\beta}$ takes values in $U(1)$, it follows that $(h_{\alpha\beta\gamma})^{n}$ is a trivial class. One can prove that for any torsion class in $H^{3}(Y, \mathbb{Z})$ there exists a corresponding twisted vector bundle \cite{AS}, therefore, if we do not fix the number of D-branes in the stack, equation \eqref{AFieldStack} can be satisfied if an only if $[\{g_{\alpha\beta\gamma}^{-1}\eta_{\alpha\beta\gamma}\}]$ is a torsion class. Hence, if we call $[h]$ the torsion class $[\{h_{\alpha\beta\gamma}\}]$ determined by the twisted bundle, the Freed-Witten anomaly cancellation for a stack of D-branes becomes \cite{Kapustin}:
\begin{equation}\label{FWStack}
	[H]\vert_{Y} + [h] = W_{3}(Y),
\end{equation}
which can be satisfied by a suitable $[h]$ if and only if $[H]\vert_{Y}$ is torsion. Therefore, if we do not impose a constraint on the number of D-brane in the stack, \emph{the only condition that the Freed-Witten anomaly imposes on a world-volume is that the $H$-flux, restricted to it, is an exact form}. In particular, if $H$ is exact on the whole-space time, there are no Freed-Witten anomalous world-volumes, even if, for some of them, there is a minimum number of D-branes that must be present in the stack, which is bigger than $1$.

Even for stack of D-branes, if the $H$-flux is not zero when restricted to the D-brane, there are large gauge transformations $B \rightarrow B + \Phi$ and $\frac{1}{n} \Tr F \rightarrow \frac{1}{n} \Tr F - \Phi$, therefore we cannot fix a gauge theory on the D-brane. When the $H$-flux is $0$ on $Y$, we can choose a representative $\{g_{\alpha\beta\gamma}, 0, 0\}$ for the $B$-field gerbe, so that equation \eqref{AFieldStack} becomes:
\begin{equation}\label{AFieldStackFlat}
	\check{\delta}^{1}\{h_{\alpha\beta}, A_{\alpha}\} = \{g_{\alpha\beta\gamma}^{-1}\eta_{\alpha\beta\gamma}, 0, \textstyle \frac{1}{n} \displaystyle \Tr F\}.
\end{equation}
This is the equation of a \emph{non-integral} vector bundle. Therefore, in the general case, when we can fix a gauge theory on a stack of D-branes, it is a gauge theory on a non-integral vector bundle, which therefore has not necessarily integral Chern classes. Of course we miss the information about the torsion part, and the Wilson loop of the $A$-field is a section of a line bundle over the loop space of $Y$, whose anomaly compensates the one of the other two terms of the world-sheet path-integral measure. We remark that, contrary to the case of a single D-brane, the flat holonomy of the $B$-field gerbe on $Y$ can be \emph{any class in $H^{2}(Y, U(1))$}, not necessarily one whose corresponding first Chern class in $H^{3}(X, \mathbb{Z})$ is $W_{3}(Y)$.

As above, if $w_{2}(Y) = 0$ and the $B$-field gerbe, restricted to $Y$, has trivial holonomy, than we can choose transition functions equal to $1$ for both the gerbes, so that \eqref{AFieldStackFlat} becomes:
\begin{equation}\label{AFieldStackStandardGT}
	\check{\delta}^{1}\{h_{\alpha\beta}, A_{\alpha}\} = \{1, 0, \textstyle \frac{1}{n} \Tr \displaystyle F\}
\end{equation}
which defines a vector bundle in the usual sense. Only in this case we recover the information even about the torsion part, and we can define the Wilson loop of the $A$-field over a closed curve as a number.

\section{Residual gauge freedom and twisted K-theory}\label{Residual}

We have said that, when $w_{2}(Y) = 0$ and the $B$-field gerbe, restricted to $Y$, has trivial holonomy, we get a gauge theory on the world-volume in the usual sense. Actually even this is not always true, unless the space-time manifold satisfies the condition $H_{1}(X, \mathbb{Z}) = 0$. We show that this fact has a connection with D-brane charge classification via twisted K-theory.

\subsection{Single D-brane}

Let us consider a space-time $X$ with one single D-brane with world-volume $Y$, supposing that $w_{2}(Y) = 0$ and the $B$-field gerbe is trivial on the whole $X$. Then the gauge theory on $Y$ should be defined on an ordinary line bundle. Nevertheless, let us suppose that this line bundle is flat. Then, the Wilson loop of the $A$-field is quantized, which means that it depends only on the homology class of the loop. Therefore, it defines an element of the cohomology group $H^{1}(Y, U(1))$. The loops we consider in order to compute the holonomy are the images of the boundary of the world-sheet $\Sigma$ under the trajectories $\phi: \Sigma \rightarrow X$, which satisfy $\phi(\partial \Sigma) \subset Y$. This means that the loop $\phi(\partial \Sigma)$, even if it is not homologically trivial on $Y$ in general, it is trivial on $X$, since $\Sigma$ itself trivializes it. Therefore, if the flat line bundle on $Y$ is the restriction of a flat line bundle on $X$, then the holonomy over any such loop will be trivial, because it coincides with the holonomy computed on $X$. Since the holonomy completely determines the connection, therefore the gauge theory, this means that \emph{the restriction on the world-volume of a flat line bundle on the whole space-time is a pure gauge}. If we have different world-volumes on $X$, each of them hosting only one D-brane, this residual gauge freedom concerns the restriction of the \emph{same} flat space-time line bundle to all of the world-volumes, since the strings starting from one D-brane and ending to another provide a way to measure the difference between the holonomies on the two world-volumes, therefore, if we fix this pure gauge for one of them, it is automatically fixed for all the others.

\subsection{Stack of D-branes}

Let us consider a space-time $X$ with one single \emph{stack} of D-branes with world-volume $Y$, still supposing that $w_{2}(Y) = 0$ and the $B$-field gerbe is trivial on the whole $X$. Then the gauge theory on $Y$ should be defined on an ordinary vector bundle. Thinking as above, we should argue that the restriction to $Y$ of a flat vector bundle on $X$ is a pure gauge. Actually this is not the case, since it must be a \emph{line} bundle anyway. In fact, if we consider $n$ different single D-branes very near to each other, by what we said above the residual gauge freedom is given only by a single space-time flat bundle, not one for each D-brane. When the D-branes get closer and closer until they coincide, we have a symmetry enhancement producing a stack of D-branes, but, up to fixing a base, we can still think of strings starting from one D-brane of the stack and ending to another, unless this choice depends on the base. Thus, we still have the possibility of fixing the gauge up to the restriction to $Y$ of a space-time line bundle. Mathematically, this is due to the following fact. Let us choose a stack of 2 D-branes, and a rank-2 complex vector bundle on $X$ which splits in the direct sum of two line bundles, with a flat connection $\mathcal{A} = A_{1} \oplus A_{2}$. Since $\Tr(AB) \neq \Tr(A)\Tr(B)$ in general for $A, B \in U(n)$, we also have in general that $\Tr(A)\Tr(A^{-1}) \neq n$: therefore, even if $\partial^{(1)}\Sigma$ and $\partial^{(2)}\Sigma$ are cohomologous in $X$, and we choose the gauge so that $\mathcal{P}e^{\int_{\partial^{(1)}\Sigma} \phi^{*}\mathcal{A}} \cdot \mathcal{P}e^{\int_{\partial^{(2)}\Sigma} \phi^{*}\mathcal{A}} = I_{n}$, it does not mean that the product of the traces is $n$ as for the trivial connection. In fact, in the present example the measure is:
	\[\begin{split}
	e^{iS} &= \cdots \Tr\, \begin{bmatrix}	e^{\int_{\partial^{(1)}\Sigma} \phi^{*}A} & 0 \\ 0 & e^{\int_{\partial^{(1)}\Sigma} \phi^{*}A'} \end{bmatrix} \cdot \Tr\, \begin{bmatrix}	e^{\int_{\partial^{(2)}\Sigma} \phi^{*}A} & 0 \\ 0 & e^{\int_{\partial^{(2)}\Sigma} \phi^{*}A'} \end{bmatrix} \\
	&= \cdots \bigl( e^{\int_{\partial^{(1)}\Sigma} \phi^{*}A} + e^{\int_{\partial^{(1)}\Sigma} \phi^{*}A'} \bigr) \bigl( e^{\int_{\partial^{(2)}\Sigma} \phi^{*}A} + e^{\int_{\partial^{(2)}\Sigma} \phi^{*}A'} \bigr)
\end{split}\]
which, expanding the product, provides four terms that, when $Y$ and $Y'$ were different, were the four possibilities for the strings. In this case, being $\partial^{(1)}\Sigma$ and $\partial^{(2)}\Sigma$ cohomologous, the result is:
	\[e^{iS} = 2 + e^{\int_{\partial^{(1)}\Sigma} \phi^{*}A + \int_{\partial^{(2)}\Sigma} \phi^{*}A'} + e^{\int_{\partial^{(1)}\Sigma} \phi^{*}A' + \int_{\partial^{(2)}\Sigma} \phi^{*}A}
\]
so that, if we fix $A$, we can find $A'$.

In general, if we have more than one stack of D-branes in $X$, the residual gauge freedom is given by a \emph{unique flat} space-time \emph{line} bundle, restricted to each world-volume.

\subsection{Twisted K-theory} From the discussion above, it follows that, when the $B$-field is trivial, the residual gauge freedom is measured by the space-time cohomology class $H^{1}(X, U(1))$. Actually $H^{1}(X, U(1)) \simeq \Hom(H_{1}(X, \mathbb{Z}), U(1))$: in fact, a flat line bundle on $X$ is given by a representation of $\pi_{1}(X)$ on $U(1)$, i.e.\ by a group homomorphism $\rho: \pi_{1}(X) \rightarrow U(1)$. Since $U(1)$ is abelian, $\rho$ uniquely projects to a representation of the abelianization of $\pi_{1}(X)$, which is $H_{1}(X, \mathbb{Z})$ \cite{Hatcher}. Therefore, the group of flat line bundles on $X$, i.e.\ $H^{1}(X, U(1))$, is canonically isomorphic to $\Hom(H_{1}(X, \mathbb{Z}), U(1))$.\footnote{This could be proven also using the universal coefficient theorem \cite{Hatcher}.} Therefore, $H^{1}(X, U(1)) = 0$ if and only if $H_{1}(X, \mathbb{Z}) = 0$, hence the residual gauge freedom is not present when the space-time manifold has vanishing first homology group.

Let us consider the case of a stack of $n$ D-branes, for $n \geq 1$, such that the $B$-field is flat on the world-volume. We consider the twisted K-theory classification of D-brane charges, which has a meaning for $[H]$ torsion of the whole $X$ and $H = 0$ as a form \cite{Witten}. The gauge bundle satisfies $h_{\alpha\beta}h_{\beta\gamma}h_{\gamma\alpha} = h_{\alpha\beta\gamma} \cdot I_{n}$, where $h_{\alpha\beta}$ takes values in $U(n)$, while $h_{\alpha\beta\gamma}$ takes values in $U(1)$ and is locally constant. In particular, the bundle defines a twisted K-theory class on $Y$, whose twisting class is $[h] = [\{h_{\alpha\beta\gamma}\}] \in \check{H}^{2}(Y, \underline{U}(1)) \simeq H^{3}(Y, Z)$, which is a torsion class being $h_{\alpha\beta\gamma}$ locally constant. Since, by equation \eqref{FWStack}, $[h] = W_{3}(Y) - [H]\vert_{Y}$, for $i: Y \rightarrow X$ the embedding of the world-volume in the space-time, being $[H]$ is torsion on the whole $X$, there is a well-defined Gysin map $i_{!}: K_{[h]}(Y) \rightarrow K_{[H]}(X)$ \cite{Karoubi}, and the image of the K-theory class $[\{h_{\alpha\beta}\}]$ under this map is the D-brane charge in twisted K-theory. The point is that $K_{[H]}(X)$ is defined only up to \emph{non-canonical} isomorphism. In fact, let us choose two representing cocycles of $[H]$, the latter though of as an element of $\check{H}^{2}(X, \underline{U}(1))$, which we call $\{g_{\alpha\beta\gamma}\}$ and $\{g'_{\alpha\beta\gamma}\}$. Then we can define the twisted K-theory groups $K_{g}(X)$ and $K_{g'}(X)$, generated by twisted bundles with twisting cocycle $g$ and $g'$ respectively: it is easy to see that any cochain $\{g_{\alpha\beta}\}$ such that $\{g'_{\alpha\beta\gamma}\} = \{g_{\alpha\beta\gamma}\} \cdot \check{\delta}^{1}\{g_{\alpha\beta}\}$ realizes an isomorphism between the two groups, but two different cochains will provide different isomorphisms. Instead, since $[H]$ is torsion and $H = 0$ as a form, let us consider the twisted K-theory group $K_{\Hol(B)}(X)$, where $\Hol(B) \in H^{2}(X, U(1))$ is the flat holonomy of the $B$-field: it is defined by \emph{non-integral bundles}, not generically twisted bundles, such that the locally constant twisting cocycle represents the class $\Hol(B)$. One can construct a well-defined Gysin map $i_{!}: K_{[h]}(Y) \rightarrow K_{\Hol(B)}(X)$, where this time $[h]$ is the twisting class $[\{h_{\alpha\beta\gamma}\}]$ of the gauge bundle in the constant sheaf $U(1)$ on $Y$. Of course, two different holonomies with the same first Chern class will produce non-canonically isomorphic groups, but we do not need to consider this. Is the group $K_{\Hol(B)}(X)$ well-defined? Actually it is the case if and only if $H^{1}(X, U(1)) = 0$, i.e.\ if and only if the residual gauge freedom vanishes. In fact, for two locally constant cocycles $\{g_{\alpha\beta}\}$ and $\{g'_{\alpha\beta}\}$ representing $\Hol(B)$, an isomorphism between $K_{g}(X)$ and $K_{g'}(X)$ is provided by a cochain $\{g_{\alpha}\}$ such that $\{g'_{\alpha\beta}\} = \{g_{\alpha\beta}\} \cdot \check{\delta}^{0}\{g_{\alpha}\}$: such a cochain is defined up to a cocycle, but, if $H^{1}(X, U(1)) = 0$, every cocycle is a coboundary, and a coboundary does not affect the isomorphism, since it changes the gauge choice for the transition functions but not the bundle, as it is easy to see. Therefore the isomorphism is canonical. This shows that the condition $H^{1}(X, U(1)) = 0$ allows for a well-defined K-theory classification group, both for the ordinary case (in which otherwise we have to quotient out by the residual gauge freedom) and the twisted case.

\section{Conclusions}\label{Conclusions}

We have seen that the gauge theory on a Freed-Witten anomaly free world-volume $Y$ is not always a gauge theory in the canonical sense, since this can happen only when $w_{2}(Y) = 0$ and the $B$-field gerbe, restricted to $Y$, has trivial holonomy. When the $B$-field is flat on $Y$, the gauge theory is defined on a non-integral vector bundle, which has well-defined Chern classes, but not belonging in general to the integral lattice in the de-Rham cohomology. Moreover, we miss the information about the torsion part, corresponding for example to the Aharonov-Bohm effect in electromagnetism. For a generic non-flat $B$-field, there are the large gauge transformations which forbid to have any canonically defined gauge theory. Moreover, when the first integral homology class of the space-time is non-vanishing, there is a residual gauge freedom represented by flat line bundles on the space-time, and the condition for this residual freedom to disappear is analogous, for flat $B$-field, to the one allowing to canonically define the D-brane charges via twisted K-theory, using non-integral bundles instead of generic twisted bundles.


\bibliographystyle{amsalpha}

\end{document}